\documentclass[12pt]{iopart}
\usepackage{amsfonts}
\usepackage{subfig}
\usepackage{graphicx,psfrag}
\begin{document}
\title[Quantum maps]{Kraus Mapping for atom-cavity and reservoir system}

\author{B. Mej\'ia, H. A.  Castillo} 

\address{Seccion Fisica, Pontificia Universidad Catolica de Per\'u, Av. Universitaria 1801, San Miguel, Lima 32, PE}
\ead{hcastil@pucp.edu.pe}

\vspace{10pt}
\begin{indented}
\item[] February 2016
\end{indented}

\begin{abstract}
We propose a way to understand the evolution of an open quantum system using a description that dispenses a continuous evolution in time, by discrete operators entangled states, in its most direct and fundamental way. We show that the successive application of these operators in very small time intervals reproduce continuous evolution. It describes and compares the temporal evolution of an open quantum system of three levels, for which the Lindblad equation is solved to obtain the density matrix function of time, a method is developed to find  Kraus operators time dependent and also finds constant Kraus operators differentials operate computationally $ n $ times evolving in discrete-time $ \Delta t = \tau = t / n $. It is seen in the example of atom cavity inside a reservoir that by calculating the distance and the relative error we define is a relationship with the evolution of discrete steps and physical variables as $ \omega $ which presuppose a natural discretization found time.
\end{abstract}

\section{Introduction}
The study of the decoherence due to the interaction between a quantum state and its environment \cite{Schlosshauer}, related to the practical use of quantum computing systems \cite{Nielsen}, is an area of greater interest both physical and technological \cite{Lidar,Xue,Pechukas}. The phenomenon of decoherence is intrinsic to the degree of entanglement of a quantum system and its interaction with the environment, which in many cases is considered as a thermal bath having many degrees of freedom, causing a loss of coherence when a measurement is independent of the state of the bath \cite{Li,Rivas} .

Entanglement as a quality of a physical system that is unique to quantum mechanics \cite{Einstein,Horodecki}, is a resource for quantum communication \cite{Ekert, Bennett}  and quantum computing \cite{Shor,Gottesman}   is often an advantage in the preparation of states with a definite, measurable and in proper condition persists in time feature. This is used as information , establishing analogues of classical states via the specific preparation of a state with a given entanglement. 
The dynamics of the quantum states used for information, and the degree of entanglement between them, are complementary problems requiring an initial evaluation of the interaction of a quantum system with a cavity in which it is contained, and the environment, measuring the degree of entanglement which modifies them. This dynamic solution, corresponding to the Lindblad equation \cite{Lindblad} for the density matrix given an initial condition, can be written as operators so as to take into account: the dissipation in the thermal bath, the information system's interaction with the cavity, and the solution of such operators' evolution of the density matrix, which is not unique.
The interaction of the atom--cavity system with its environment that we have analyzed has an analytical solution that can be written as Kraus operators \cite{kraus} not developed in other works in detail. There are more than one group of Kraus operators that solve the problem because there is some freedom in how we choose the matrix elements in the chosen representation, which shows that decoherence does not occur equally in every part of the entanglement which is measured in the system. This allows selecting possible relationships between the initial states with comparable coherence times with the possibility of making operations or changes between them. Moreover, the measurement problem persists in the Hilbert space that corresponds to the states containing the information. Also, there are operations that can be simulated in an enlarged space \cite{Dicandia}, in this case an arrangement is given extra dimensions, for the Kraus operators.

\section{A system of an atom--cavity in a  reservoir}
We consider the atom--cavity system in the microwave regime, in contact with a reservoir at very low temperature  ($ T \rightarrow 0 $). This allows direct measurements of the effect of the atom--cavity and the cavity--environment couplings. In such a situation, it is possible to analyze the time evolution of a specific state using the Lindbland equation (the semigroup master equation) \cite {Lindblad}. This can be solved analytically in this system, using its explicit form (for $ \hbar = 1 $)  \cite{Haroche,Rosario}

\begin{equation}
\frac{d\rho(t)}{dt}=-i[H_{a}+H_{c}+H_{ac},\rho(t)]+\kappa[2a\rho(t)a^{\dag}-a^{\dag}a\rho(t)-\rho(t)a^{\dag}a] \label{Lic}
\end{equation}
where $ H_ {a} = \frac {1} {2} \omega_{eg} \sigma_ {z} $,  $ H_{c} = \omega_{c}  a^ {\dag} a $, and $H_{ac}=\frac{-i\Omega}{2}(a\sigma_{+}-a^{\dag}\sigma_{-})$  are the Hamiltonian of the atom, cavity, and atom--cavity interaction, respectively. $ \kappa $ is a decay constant that is related to the quality factor of the cavity $ Q $, and $ \kappa = \omega_{c} / Q $.
All other values
$ \sigma_{+} \equiv | e \rangle \langle g | $, $ \sigma_{-} \equiv | g \rangle \langle e | $ are  \emph{up} and \emph{down} operators of the atom and $ \sigma_ {z} \equiv | e \rangle \langle e | - | g  \rangle \langle g | $ is related to the spin flip operator in the Pauli algebra of  matrices.

In  the above equation, we consider the resonant case, $ \omega_{eg} = \omega_{c} = \omega $
and initially  an empty state for the cavity.  Then there is at most one state of excitation in the atom--cavity system. The Hilbert space is given by three vectors, which we denote by 
$|1\rangle\equiv|e0\rangle$, $|2\rangle\equiv|g1\rangle$, $|3\rangle\equiv|g0\rangle$
\begin{figure}[!ht]
 \centering
\subfloat[$|1\rangle\equiv|e0\rangle$]{\label{fig:gull}\includegraphics[width=0.3\textwidth]{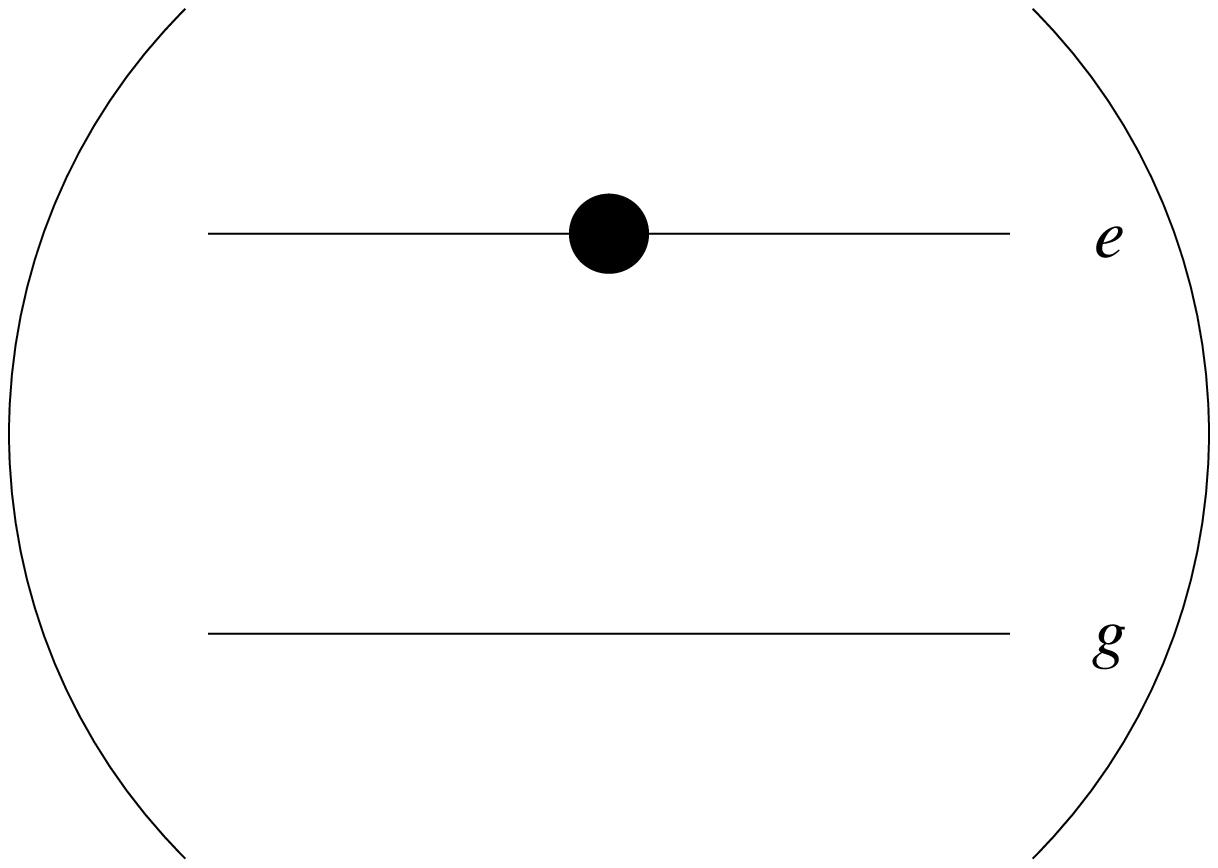}}
  ~ 
  \subfloat[$|2\rangle\equiv|g1\rangle$]{\label{fig:tiger}\includegraphics[width=0.3\textwidth]{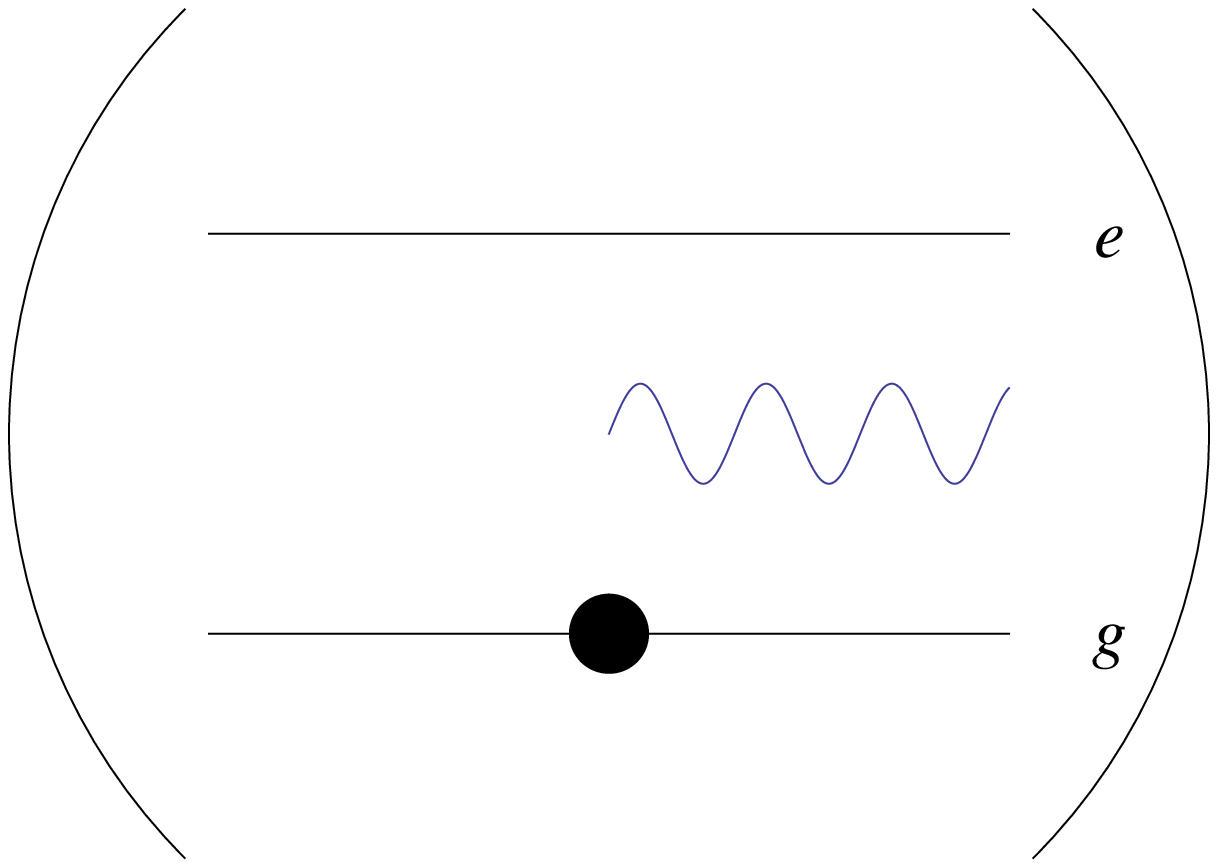}}
  ~ 
  \subfloat[$|3\rangle\equiv|g0\rangle$]{\label{fig:mouse}\includegraphics[width=0.3\textwidth]{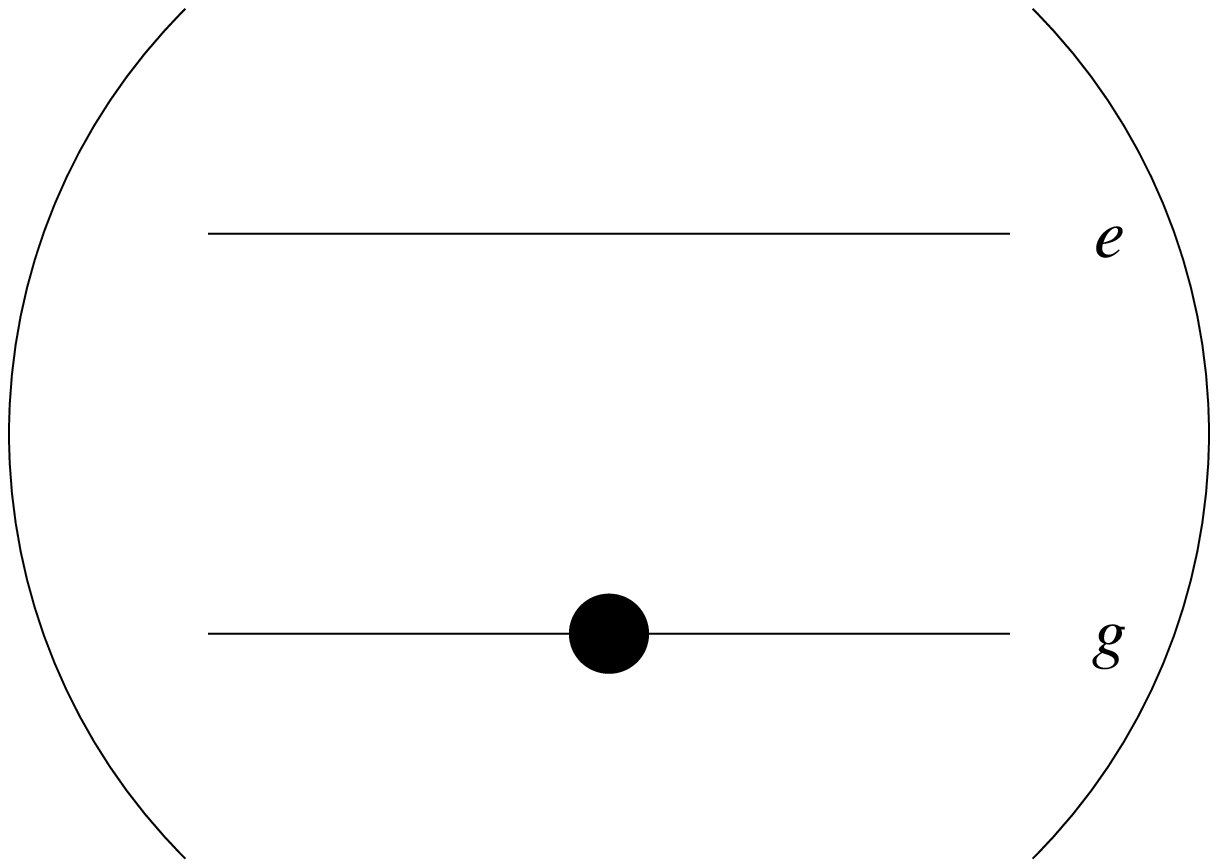}}
  \caption{(a), (b) and (c) are  representations for the three basis vectors }
  \label{fig:vectores base}
\end{figure}

Here,  
$|e\rangle$ is the excited state of the atom, $ | g \rangle $ is its ground state, $ | 1 \rangle $  is the presence  of photons in the cavity, and $ | 0 \rangle $ otherwise.

Eq. (\ref {Lic}) can be developed  into the differential equations
\numparts
\begin{eqnarray}
\frac{d\rho _{11}(t)}{dt}&=-\frac{1}{2} \Omega  \left(\rho _{12}(t)+\rho _{21}(t)\right)\\
\frac{d\rho _{22}(t)}{dt}&=\frac{1}{2} \Omega  \left(\rho _{12}(t)+\rho _{21}(t)\right)-\kappa  \rho_{22}(t)\\
\frac{d\rho _{12}(t)}{dt}&=\frac{1}{2} \Omega  \left(\rho _{11}(t)-\rho _{22}(t)\right)-\frac{1}{2}
   \kappa  \rho _{12}(t)\\
\frac{d\rho _{21}(t)}{dt}&=\frac{1}{2} \Omega  \left(\rho _{11}(t)-\rho _{22}(t)\right)-\frac{1}{2}
   \kappa  \rho _{21}(t)\\
\frac{d\rho _{33}(t)}{dt}&=\kappa \rho _{22}(t)\\
\frac{d\rho _{13}(t)}{dt}&=-\frac{1}{2} \Omega  \rho _{23}(t)-i \omega  \rho _{13}(t)\\
\frac{d\rho _{23}(t)}{dt}&=\frac{1}{2} \Omega  \rho _{13}(t)+\left(-\frac{\kappa }{2}-i \omega
   \right) \rho _{23}(t)\\
\frac{d\rho _{31}(t)}{dt}&=-\frac{1}{2} \Omega  \rho _{32}(t)+i \omega  \rho _{31}(t)\\
\frac{d\rho _{32}(t)}{dt}&=\frac{1}{2} \Omega  \rho _{31}(t)+\left(-\frac{\kappa }{2}+i \omega
   \right) \rho _{32}(t)
\end{eqnarray} \label{9Eq}
\endnumparts
describing the density matrix components $ \rho_ {kl} (t) $ then we find the solutions for $ \rho_{kl}(t) $, where $ k $, $ l $, $ m $ and $ n $ takes the values $1$, $ 2 $ or $ 3 $, which satisfy the relation
\begin{equation}
\rho_{kl}(t)=\sum_{mn}F^{mn}_{kl}(t) \rho_{mn}(0) \label{EQF}
\end{equation}
The resolution process is changing in vector form the density matrix  as follows:
\begin{equation}
\dot {\left[\rho(t)\right]}=A \left[\rho(t)\right]\Rightarrow \left[\rho(t)\right]= e^{A t}\left[\rho(0)\right]=F(t) \left[\rho(0)\right]
\end{equation}
where \begin{small}
$ \left[ \rho(t)\right] \equiv
\left(
\begin{array}{c c c c c c c c c }
 \rho _{11}(t) & \rho _{12}(t) & \rho _{13}(t) &  \rho _{21}(t) &  \rho _{22}(t) &
 \rho _{23}(t) &  \rho _{31}(t) & \rho _{32}(t) & \rho _{33}(t)
\end{array}
\right)^{T}
$
\end{small} and  $A$ is the matrix extract to Eq. (\ref{9Eq}),  $F(t)= e^{A t}$

The functions $F^{mn}_{kl} (t) $ are elements of the matrix $F(t)$  and the accompanying values $\rho_{mn}(0) $ are the elements of the initial density matrix.
The functions  $F^{mn}_{kl} (t) $ can also be determined from three \emph{generating} functions ($\Lambda_{+}(t)$,$\Lambda_{-}(t)$, $\Lambda_{0}(t)$) and $e^{i \omega t }$:
\begin{equation}
\Lambda_\pm(t)\equiv g(t) \frac {\sinh\left(\frac { \Omega \sqrt{\gamma^{2}-1}}{2}  t \pm \phi\right)}{ \sqrt{\gamma^{2}-1}} ,\quad \Lambda_{0}(t)\equiv g(t) \frac {\sinh\left(\frac { \Omega \sqrt{\gamma^{2}-1}}{2}  t \right)}{ \sqrt{\gamma^{2}-1}}
\end{equation}

where  $\gamma\equiv\frac{\kappa}{2\Omega}=\cosh(\phi)$ and $g(t) = e^{-\kappa t/4}$.  When the system is mostly dissipative, $\gamma>1$  and $ \Omega>0 $ the argument of these generating functions is real and decreases. Otherwise $ \gamma <1$ and $ \kappa> 0$ have oscillating behavior and decrease. 

Now if we want to express the complete solution of the density matrix $ \rho (t) $ in its spectral form, it would be
\begin{equation}
\rho(t)=\sum_{kl}\rho_{kl}(t)|k\rangle\langle l| \label{erho}
\end{equation}
If we substitute in (\ref{erho}) the Eq. (\ref{EQF}), we have
\begin{equation}
\rho(t)=\sum_{kl}\left(\sum_{mn}F^{mn}_{kl}(t) \rho_{mn}(0) \right)|k\rangle\langle l| \label{eqEF0}
\end{equation}
After some calculations, in order to arrange conveniently the last equation, we have\begin{equation}
\rho(t)=\sum_{mn}\left( \sum_{kl}F^{mn}_{kl}(t)|k\rangle\langle l|\right) \rho_{mn}(0) \label{eqEF1}
\end{equation}
Here what is inside the parentheses should be noted: it contains the evolution of the system which is independent of the initial conditions of $ \rho $. This is very useful when analyzing decoherence, because the initial matrix elements can continually change at our convenience.

\section{The Kraus dynamic mapping for the evolution of an atom--cavity system}
From Eq. (\ref {eqEF1}), the proposed mapping $ \varphi $ should take the form of a sum operator \cite{kraus}
\begin{equation}
\rho(t)= \varphi \left( \rho(0)\right) =\sum_{\mu}K_{\mu}(t)\rho(0)K^{\dagger}_{\mu}(t) \label{rhot}
\end{equation}
This means that, for this other way to express $ \rho(t) $, we have operators $ K_{\mu}(t) $ that enable the evolution of the matrix $ \rho(0) $ up to a time $ t$, where $ N $ is the dimension of the Hilbert space and  $ \mu$ is a value up to $ N^2-1$ \cite{Haroche}.
Now if we write the spectral expression of  the $ \rho(0) $, we have
\begin{equation}
\rho(0)=\sum_{mn}\rho_{mn}(0)|m\rangle\langle n| \label{Erho0}
\end{equation}
If now we place $ \rho(0)$ of the last equation into (\ref{rhot}):
\begin{equation}
\rho(t)=\sum_{\mu}K_{\mu}(t)\left(\sum_{mn}\rho_{mn}(0)|m\rangle\langle n|\right) K^{\dagger}_{\mu}(t) \label{rho1t}
\end{equation}
This, after sorting in the most convenient way to compare with Eq. (\ref {eqEF1}), gives us
\begin{equation}
\rho(t)=\sum_{mn}\left( \sum_{\mu}K_{\mu}(t)|m\rangle\langle n|K^{\dagger}_{\mu}(t)\right)  \rho_{mn}(0) \label{eqEK}
\end{equation}
This form will allow us to determine the evolution of each element $ |m \rangle \langle n| $ using the  Kraus operators. 
If now we match Equations (\ref{eqEK}) and (\ref{eqEF1}), since both are $ \rho(t) $, we have
\begin{equation}
\sum_{mn}\left( \sum_{kl}F^{kl}_{mn}(t)|k\rangle\langle l|\right)  \rho_{mn}(0) =\sum_{mn}\left( \sum_{\mu}K_{\mu}(t)|m\rangle\langle n|K^{\dagger}_{\mu}(t)\right)  \rho_{mn}(0) \label{eqFK}
\end{equation}

From this last equation, we can see that every element of $ \rho (t) $ evolves in such a way as to satisfy
\begin{equation}
\sum_{kl}F^{kl}_{mn}(t)|k\rangle\langle l| =\sum_{\mu}K_{\mu}(t)|m\rangle\langle n|K^{\dagger}_{\mu}(t) \label{eqFK0}
\end{equation}

\begin{figure}[!ht]
 \centering
\includegraphics[scale=0.8]{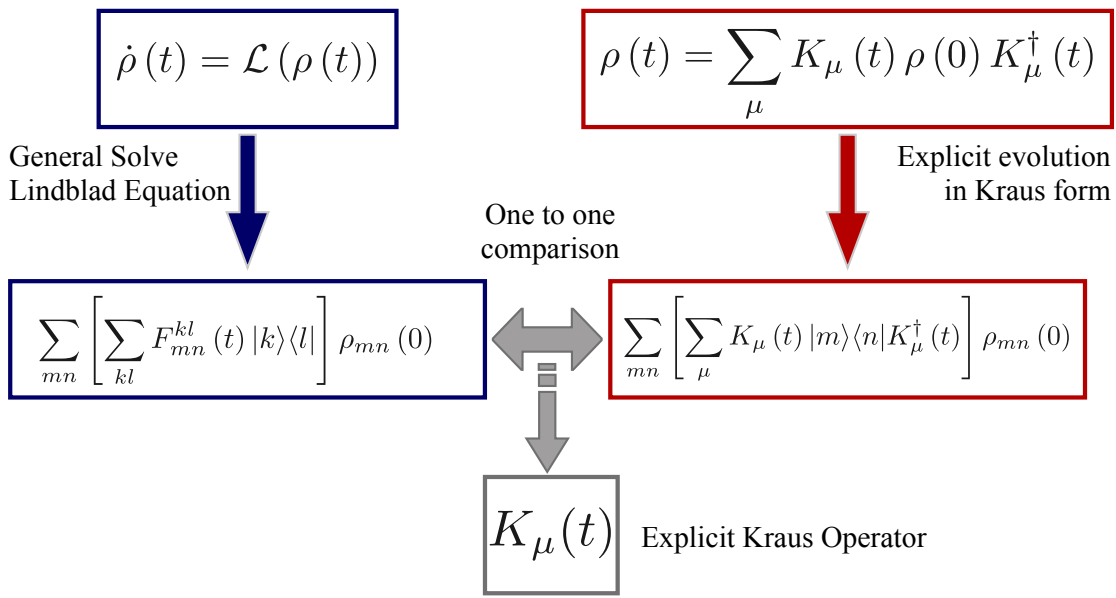} 
  \caption{In the diagram $\mu$  takes values between 0 and $\mu_{max}$. The value  $\mu_{max}  = 0,1,2 ...$ is chosen for adequate comparison.  }
  \label{fig:diagram}
\end{figure}

\subsection{The approximation of the Kraus operators and the explicit determination of the operators $K_{\mu}(t)$}

We define the operators $\widetilde{K}_{\mu}$ \cite{Haroche} as 
\begin{equation}
\widetilde{K}_{\mu}=\left\lbrace 
\begin{array}{cc}
\mathbb{I}+A \tau  & \mu=0 \\
  L_\mu \sqrt{\tau}& \mu>0 \\
\end{array}\right. \label{akraus}
\end{equation}

In the particular case under consideration, $ A $ and $ L_\mu $ are $ 3 \times 3 $ matrices. This definition of $ \widetilde {K}_{\mu} $ is required to obtain the Lindblad equations in an analytical way. For small values of $ \tau $, expanding to first order, and ordering the result of adding in a convenient manner, yields
\begin{equation}
\widetilde{\rho}(\tau)=\sum\widetilde{K}_{\mu} \rho(0)\widetilde{K}_{\mu}^\dagger\approx\rho(0)+\tau \left(  A \rho(0)+ \rho(0) A^\dagger+ \sum_\mu L_\mu \rho(0) L_{\mu}^\dagger \label{edrho} \right) 
\end{equation} 

This can be compared  with the complete solution for the density matrix $ \rho (t) $, Eq. (\ref{rhot}), for small times $ \tau $, obtaining a first approximation

\begin{equation}
\widetilde{K}_0=\mathbb{I}+A \tau=\left(
\begin{array}{ccc}
 1 & 0 & 0 \\
0& 1 & 0   \\
 0 & 0 & 1 \\
\end{array}
\right)+ \left(
\begin{array}{ccc}
0 & -\frac{ \Omega }{2} & 0 \\
 \frac{ \Omega }{2} &  -\frac{\kappa }{2} & 0
   \\
 0 & 0 & i \omega \\
\end{array}
\right)\tau \label{aktau0}
\end{equation}

\begin{equation}
\widetilde{K}_{\mu>0}=L_\mu \sqrt{\tau}=\left(
\begin{array}{ccc}
 0 & 0 & 0 \\
 0 & 0 & 0 \\
 0 &  \ell_{\mu} & 0 \\
\end{array}
\right)\sqrt{\tau}\label{aktaumu}
\end{equation}
This must satisfy $\sum |\ell_{\mu}|^ 2 = \kappa $. In this way we can get the form of the Kraus operators for small times $ \tau $. \\
Now, if we want to determine the Kraus operators  $ K_{\mu}(t) $ in general, we must determine the functions $ F^{mn}_{kl}(t) $ in Equation (\ref {eqFK0}). As the operator $ K_{\mu} (t) $ \cite{Haroche} is part of a summation over the  subscript $ \mu $ running from $ 0 $ to the maximum dimension of the Hilbert space, which is $ 3^2-1 $, we must test in which of the cases would Equation (\ref{eqFK0}) be fulfilled. 
Here, each $ K_{\mu_{ij}} (t) $ is an element of the time dependent array  $ K_{\mu} (t) $.
As $\rho(t)$ was obtained in terms of $ F^{mn}_{kl}$, and as it is a function of $ e^{i\omega t} $ and $ e^{-i\omega t} $, which are complex functions, but the generating functions $ \Lambda_{+} (t) $, $ \Lambda_{-} (t) $ and $ \Lambda_{0} (t) $ are real functions (see Section 2), we can find a relationship that should exist with each element of $\rho_{kl} (t) $ and $ K_{\mu_{ij}} (t) $ that also fulfills the condition of renormalization, i.e.:
\begin{equation}
\sum _{\mu=0}^{N^2-1}  K_\mu^{\dagger }K_\mu=\mathbb{I} \label{ckraus}
\end{equation}
Here $N$ is the number of dimensions of the system, so in our case the sum would reach a maximum of nine summands including zero.

If we want to analyze cases of finding Kraus operators with different summands we obtain:
\begin{itemize}
\item {Case $\rho(t)=K_{0}(t) \rho(0)K_{0}^{\dagger }(t)$ ($\mu=0$)}\\
By comparing the elements of the matrix $ K_ {0} (t) $ satisfying Equation (\ref{eqFK0}) it can be seen that no solutions are possible because there are more equations than unknowns in its development.
\item{Case $ \rho(t) = \sum_{\mu = 0}^{1} K_{\mu} (t) \rho(0) K_{\mu}^{\dagger} (t) $ ($ \mu = 0,1$)} \\
By comparing the elements of the matrix $ K_{0} (t) $ and $ K_{1} (t) $ satisfying Equation (\ref {eqFK0}), it can be seen that there are only particular solutions when the generating functions take on trivial values, which we can express in the following table:

\begin{table}[!htbp]
	\centering
\begin{tabular}[b]{|c|c|c|}
\hline
For &$\Omega\rightarrow 0$ & $\kappa \rightarrow 0$  \\ \hline
Generating functions & $\begin{array}{c}
\Lambda _-=-e^{-\frac{\kappa  t}{2}} \\ \Lambda_{0}=0\\ \Lambda _+=1
\end{array}$ & $\begin{array}{c}
\Lambda _-=-\cos
   \left(\Omega t /2\right) \\ \Lambda_{0}=\sin \left(\Omega t /2\right)\\ \Lambda _+=\cos \left(\Omega t /2\right)
\end{array}$ \\ \hline	
		\end{tabular}
	\caption { In the case of $ \Omega = 0$ there is no interaction between the atom and the cavity (they evolve independently), while for the case of $ \kappa = 0 $, there is no interaction between the atom--cavity system and the environment}\label{tab1}
\end{table}
\item{Case $\rho(t)=\sum_{\mu=0}^{2}K_{\mu}(t)\rho(0)K_{\mu}^{\dagger}(t)$ ($\mu=0,1,2$)} \\
As in the previous cases, to compare the elements of the matrix $ K_ {0} (t) $ and $ K_ {1} (t) $ satisfying Equation (\ref {eqFK0}), we obtain three equations with four unknowns. If one of the unknowns is held fixed, we can say there are different Kraus operators (the solution to the defining equation of the Kraus operators). A example could also be
\begin{equation} 
 K_{0}=\left(
\begin{array}{ccc}
1 & 0 & 0 \\
0 & 1 & 0 \\
0 & 0 & 1  \\
\end{array}
\right)+ \left(
\begin{array}{ccc}
 \Lambda _+ -1 & -\Lambda_{0} & 0 \\
 \Lambda_{0} & -\Lambda _- -1& 0 \\
 0 & 0 & e^{i \omega t }-1 \\
\end{array}
\right)\label{kf0}
\end{equation}
\begin{equation}
K_{1}=\left(
\begin{array}{ccc}
 0 & 0 & 0 \\
 0 & 0 & 0 \\
 -\frac{\sqrt{\lambda _+}}{\sqrt{2}} & \frac{-2 \gamma \Lambda_{0}^2
  +\sqrt{\lambda_+ \lambda_{-}-4 \gamma^2\Lambda_{0}^4 }}{
   \sqrt{2 \lambda _+}}  & 0 \\
\end{array}
\right)\label{kf1}
\end{equation}
\begin{equation}
K_{2}=\left(
\begin{array}{ccc}
 0 & 0 & 0 \\
 0 & 0 & 0 \\
 -\frac{\sqrt{\lambda _+}}{\sqrt{2}} & \frac{-2 \gamma \Lambda_{0}^2
  -\sqrt{\lambda_+ \lambda_{-}-4 \gamma^2\Lambda_{0}^4 }}{
   \sqrt{2 \lambda _+}} & 0 \\
\end{array}
\right)\label{kf2}
\end{equation}
where $2\gamma  \Lambda_{0}=\left(\Lambda _-+\Lambda _+\right)$ and $\lambda_{\pm}=1-\Lambda_{0}^2-\Lambda _\pm^2 $. Finally, once all the elements of the matrices $ K_{\mu} $ that satisfy the preconditions are determined, we have a tool for developing a better analysis of the system.
\end{itemize}
On the other hand, since $ K _ {\mu} (t) $ in the last case is given for any time, we can identify the value of $ \ell _ {\mu} $ of Equation (\ref {aktau0}): it would be in this case $ \ell_ {1} = - \frac {\left (\sqrt {3} -1 \right) \sqrt {\kappa}} {2 \sqrt {2}} $ and $ \ell_ {2} = - \frac {\left (\sqrt {3} +1 \right) \sqrt {\kappa}} {2 \sqrt {2}} $.\\
Since $ \tau $ is sufficiently  small, we can repeat Equation (\ref {edrho}) several times with increasing values so that we have $ \rho (t) \approx \widetilde {\rho} (n \tau) $, where $ n \gg  1$.
Since the value of $ \tau $ physically is not a differential but rather is a small value of time, we determine a range of values using the norm of the matrix, both the density matrix and each Kraus operator. This will allow fixing the values of $ \tau $ that would approximate to a measurable value.

With what has been obtained up to this point, we can say that when we solve the Lindblad equation, Eq. (\ref{Lic}), we treat it like a differential, i.e. we replace $ \frac {\Delta \rho} {\Delta t} \rightarrow \frac {d \rho} {dt} $, therefore the time is considered as an infinitesimal value, but physically,  the evolution would be discrete for $ \Delta t \ll \omega ^ {- 1} $. The discrete evolution is expressed by Eq. (\ref {edrho}) where $ \tau = \Delta t =t/n$. In the next section, we analyze and compare the continuous evolution of Eq. (\ref {rhot}) and the discrete evolution of Eq. (\ref {edrho}), several studies indicate the presence of a discrete time \cite{Page, Giovannetti}.

\section{Graphical analysis}
\begin{figure}[ht!]
   \centering
   \subfloat[]{

        \label{fig1:pr1:a}         
        \includegraphics[width=0.45\textwidth]{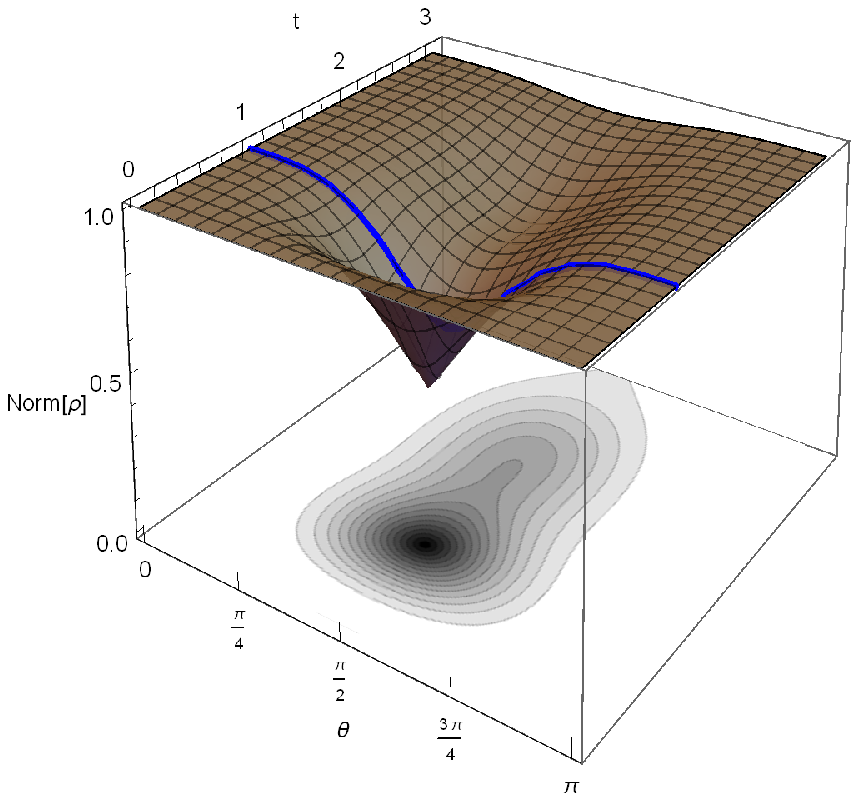}}
   \hspace{0.005\linewidth}
   \subfloat[]{ 

        \label{fig1:pr1:b}         
        \includegraphics[width=0.45\textwidth]{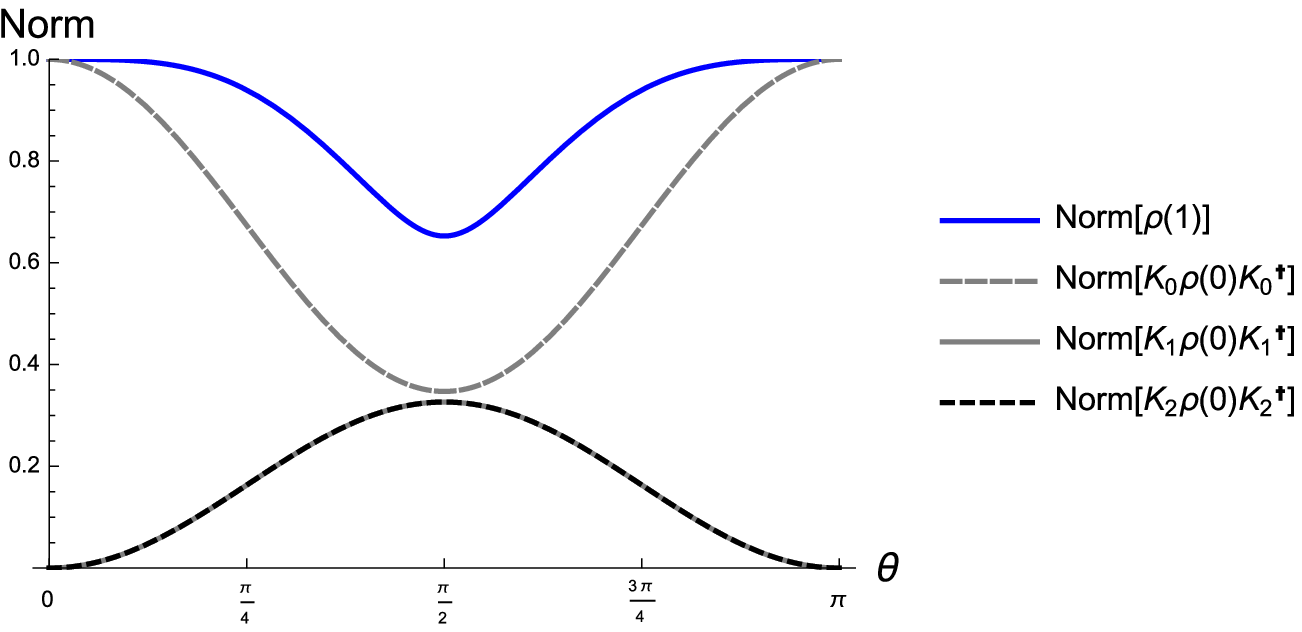}}\\
      \caption{ Plot for  $\kappa=2$, $\Omega=4$ and $\omega=2$,  .  (a) Tridimensional  plot for $\rho(t)$, for different initial conditions $\theta\in[0,\pi]$ and different time $t \in[0,3]$. (b) Particular case, Norm for $\rho (1)$ (blue) and norm for  action in individual operators$ K_0$, $K_1$ and $K_2$ on $ \rho_(0)$ as a function of $\theta$}
   \label{fig1:pr1}                
\end{figure}

\begin{figure}[ht!]
   \centering
   \subfloat[]{

        \label{fig2:pr2:a}         
        \includegraphics[width=0.48\textwidth]{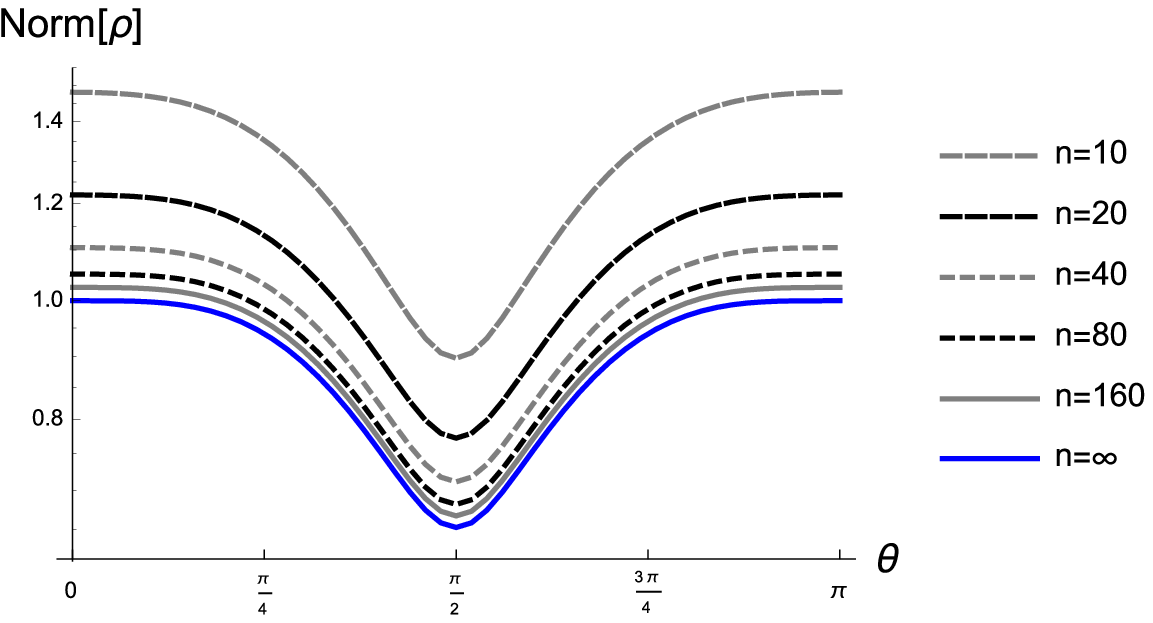}}
   \hspace{0.005\linewidth}
   \subfloat[]{ 

        \label{fig2:pr2:b}         
        \includegraphics[width=0.48\textwidth]{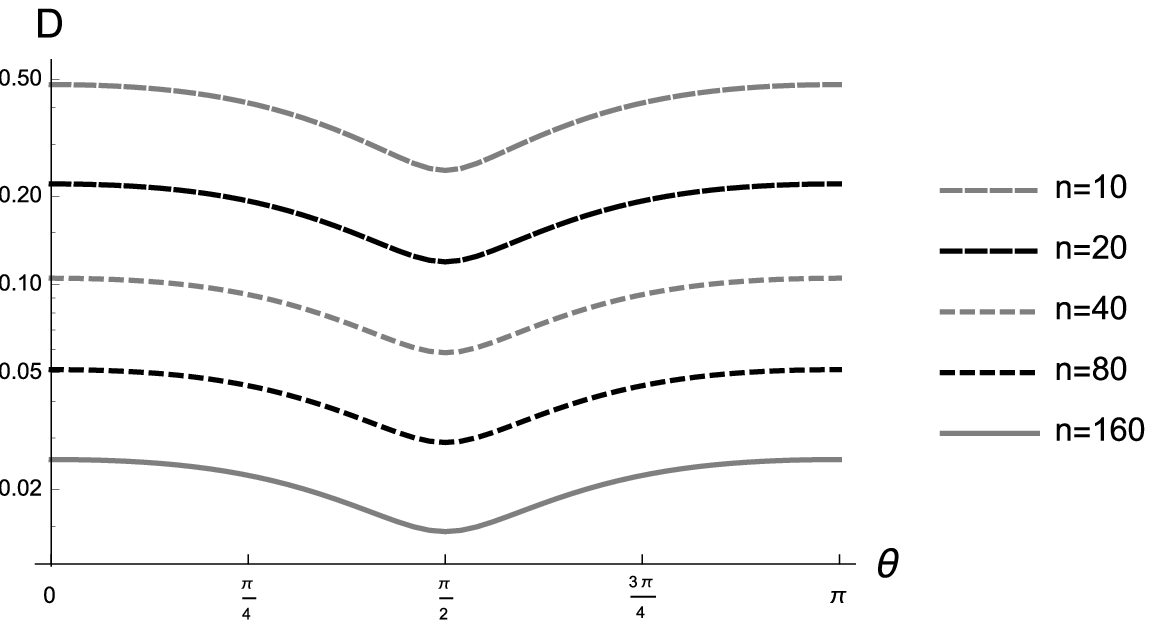}}\\
      \caption{To $\rho_ (1)$ and the same conditions of Figure \ref{fig1:pr1} and different initial conditions $\theta\in[0,\pi]$ (a)  The numerical norm  $\widetilde{\rho}(1)$ that depend of $n$.  The analytical solution $\rho(1)$ (in blue line) is indicated by $n=\infty$. (b) The Distance ($D$)  that depend of $n$ }
   \label{fig2:pr2}                
\end{figure}

For this analysis we take as the state $\vert\psi(0)\rangle=\left(  \cos \theta \vert\phi_A\rangle+\sin  \theta \vert\phi_A\rangle\right)\otimes\vert 0 \rangle$.  Then the corresponding initial density matrix  used  in Section 2 (Fig. \ref{fig:vectores base}) would be
\begin{equation} \fl
\rho(0)=\vert\psi(0)\rangle\langle\psi(0)\vert= \cos^2 \theta \vert 3 \rangle\langle 3\vert+\sin^2  \theta \vert 1\rangle\langle 1\vert+\sin  \theta \cos  \theta \vert 3\rangle\langle 1\vert+\sin  \theta \cos  \theta \vert 1\rangle\langle 3\vert
\end{equation}
Therefore it will be in matrix form:
\begin{equation}
\rho(0)=\left(
\begin{array}{ccc}
 \sin ^2\theta & 0 & \cos \theta \sin
   \theta \\
 0 & 0 & 0 \\
 \cos \theta \sin \theta & 0 & \cos
   ^2\theta \\
\end{array}
\right) 
\end{equation}
Now we use this initial  density matrix to compare the density matrices of a continuous evolution
(using the explicit formula for the matrices $K_{\mu}$) and the discrete approximation for small increments of time $\tau$, given by Equations   (\ref{rhot}) and (\ref{edrho}), respectively. We define the norm of a matrix to be its maximum singular value. 
Figure (\ref{fig1:pr1:a})  is  matrix norm for  the values of  $\kappa=2$, $\Omega=4$ and $\omega=2$ as a function of $\theta$ differents as the initial condition of the state and time $t \in[0,3]$, the line blue is show too in the figures (\ref{fig1:pr1:b})  where are the  different norm of part the Kraus operator here the contribution of each individual system evolution oprerador $K_i$ shown. Note that its importance in the evolution of the total. In the figure (\ref{fig2:pr2:a}),  the matrix norms of  $\widetilde{\rho}(1)$ (approximative) and $\rho(1)$ (analytic) are shown, it can be seen that for values greater than $n$ (the number of discrete steps) it approaches the solution given by the analytical value.  
Figures (\ref{fig2:pr2:b}) is the distance,  $D=Norm[\widetilde{\rho}(1)-\rho(1)]$  , defined as the norm of the matrix of the difference between  $\rho(t)$ analytic and $\widetilde{\rho}(n \tau)$ numerical.  

\begin{figure}[ht!]
   \centering
   \subfloat[]{

        \label{fig3:pr3:a}         
        \includegraphics[width=0.48\textwidth]{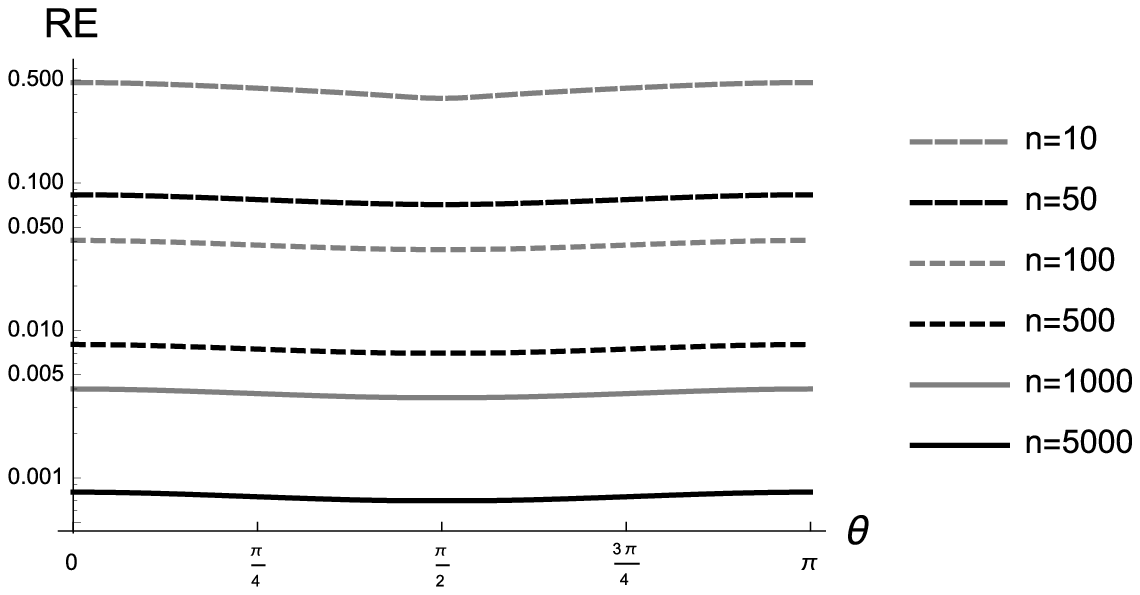}}
   \hspace{0.005\linewidth}
   \subfloat[]{ 

        \label{fig3:pr3:b}         
        \includegraphics[width=0.48\textwidth]{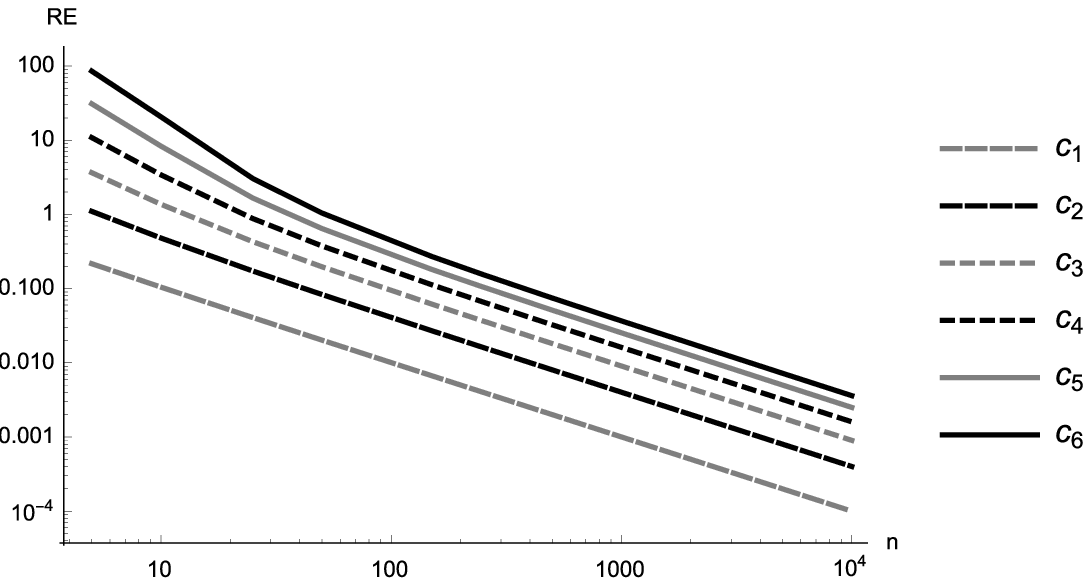}}\\
      \caption{(a) In  logarithmic scale, the Relative Error  ($RE$) and  different initial conditions $\theta\in[0,\pi]$ ,  for   $\kappa=2$, $\Omega=4$ and $\omega=2$ , where $n$ means the number of steps in the iterations, 
    (b)    In logarithmic scale the Relative Error and $n$, the number of steps in the iterations for diferent value of $\kappa$, $\Omega$ and $\omega$. The subscript  indicate the $\omega$ value ($c_{\omega}$) }
   \label{fig3:pr3}                
\end{figure}

Figure  (\ref{fig3:pr3:a}) shows the Relative Error, $RE=\frac{Norm[\widetilde{\rho}(t)]-Norm[\rho(n\tau)]}{Norm[\rho(t)]}$ versus $\theta$ . Figure  (\ref{fig3:pr3:b}) shows the Relative Error versus $n$ that means the number of divisions of time and  the number of steps in the iterations.  

\begin{figure}[ht!]
   \centering
   \subfloat[]{

        \label{fig4:pr4:a}         
        \includegraphics[width=0.48\textwidth]{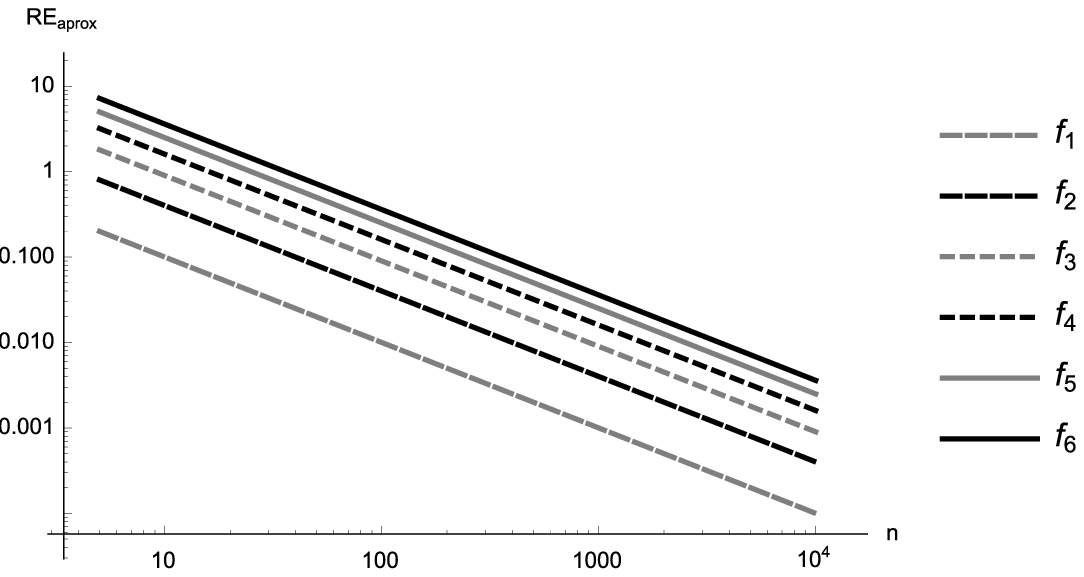}}
   \hspace{0.005\linewidth}
   \subfloat[]{ 

        \label{fig4:pr4:b}         
        \includegraphics[width=0.48\textwidth]{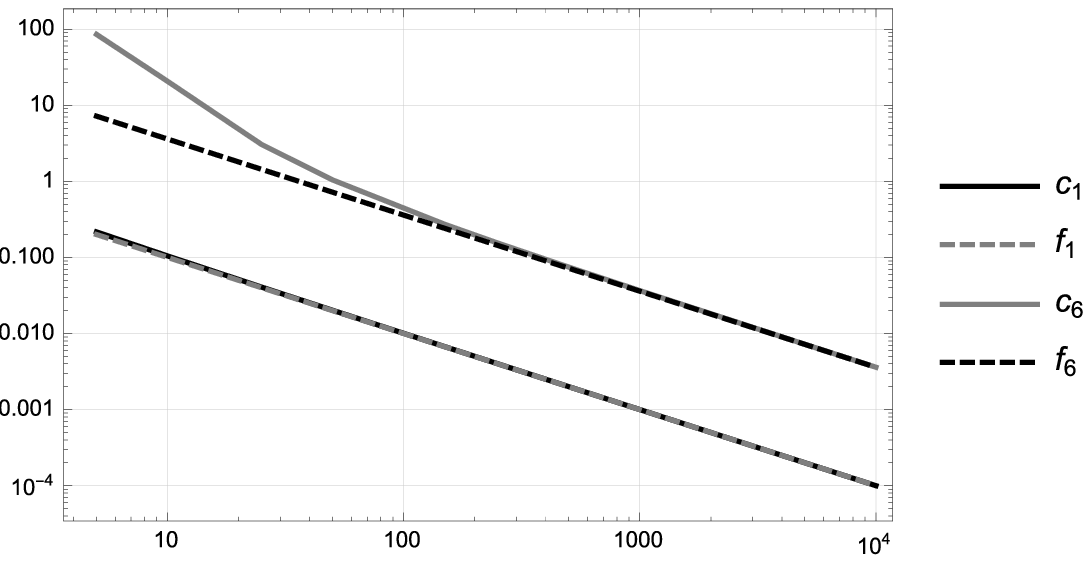}}\\
      \caption{(a)    In logarithmic scale the relative error calculate in the aproximation $RE_{aprox}=\frac{\omega^2}{n}$ and the number of steps in the iterations for diferent value of $\kappa$, $\Omega$ and $\omega$. The subscript  indicate the $\omega$ value ($f_{\omega}$) 
    (b)  In logarithmic scale the comparation to $RE$ and $RE_{aprox}$  }
   \label{fig4:pr4}                
\end{figure}

Figure  (\ref{fig4:pr4:a}) shows the Relative Error aproximation, $RE_{aprox}=\frac{\omega^2}{n}$,   versus $n$  and Figure  (\ref{fig4:pr4:b}) shows  the comparation to $RE$ and $RE_{aprox}$ versus $n$, this  allows us to control the system within a physically acceptable discrete evolutionary process because relation the $RE$ with the number of steps in the iterations.

\section{Conclusions}

The matrices $ K $ (Kraus matrices) explicitly allowing the control of all the variables in the studied system were found. In the case   $\rho(t)=K_{0}(t)\rho(0)K_{0}^{\dagger }(t)$  there is no solution for a single Kraus operator. In the case $\rho(t)=\sum_{\mu=0}^{1}K_{\mu}(t)\rho(0)K_{\mu}^{\dagger}(t)$ only nontrivial solutions of an uncoupled system or a closed system are obtained. In the case $\rho(t)=\sum_{\mu=0}^{2}K_{\mu}(t)\rho(0)K_{\mu}^{\dagger}(t)$ there exist several complete solutions (available only for $\mu \geq 2$), which allows at least one variable that can control more  the other variables and therefore there are additional solutions for new interpretations and conclusions.  The explicit development of the Kraus matrices has led to understanding that their equations are compatible with the Kraus condition of Eq.  (\ref{ckraus}).

The studied system gives us the basis for studying, with greater detail, mathematical developments using Kraus operators. Having an analytical form for the Kraus operators has led us to compare the number of steps in a numerical solution that can be valid since there is a controllable variation of the steps, this is a unique feature of the solution. Although the system is forced numerically to evolve in discrete steps, you can search for an equal value that is compatible with the physical variables ($\omega$, $\Omega$, $\kappa$) to occur in a manner acceptable analytical development, establishing a relationship between the solution by Kraus operators and approximate semigroup Linbland via the equation. This relationship between $\tau$ and $\omega$ is directly in the system in which the system saves the information in the atomic degrees of freedom. A relationship was developed between  norm Relative  Error depends on steps number,  for  control time discretization.

\section{Acknowledgments}
The authors acknowledge support from the PAIP project; the Departamento de Ciencias PUCP, and a PRECIOSA fellowship.

\section{Appendix: A}
 \begin{small}
 $ \left[ \rho(t)\right] \equiv
\left(
\begin{array}{c}
 \rho _{11}(t) \\
 \rho _{12}(t) \\
 \rho _{13}(t) \\
 \rho _{21}(t) \\
 \rho _{22}(t) \\
 \rho _{23}(t) \\
 \rho _{31}(t) \\
 \rho _{32}(t) \\
 \rho _{33}(t)
\end{array}
\right)
$,  $A=
\left(
\begin{array}{ccccccccc}
 0 & -\frac{\Omega }{2} & 0 & -\frac{\Omega }{2} & 0 & 0 & 0 & 0 & 0 \\
 \frac{\Omega }{2} & -\frac{\kappa }{2} & 0 & 0 & -\frac{\Omega }{2} & 0 & 0 & 0 & 0 \\
 0 & 0 & -i \omega  & 0 & 0 & -\frac{\Omega }{2} & 0 & 0 & 0 \\
 \frac{\Omega }{2} & 0 & 0 & -\frac{\kappa }{2} & -\frac{\Omega }{2} & 0 & 0 & 0 & 0 \\
 0 & \frac{\Omega }{2} & 0 & \frac{\Omega }{2} & -\kappa  & 0 & 0 & 0 & 0 \\
 0 & 0 & \frac{\Omega }{2} & 0 & 0 & -\frac{\kappa }{2}-i \omega  & 0 & 0 & 0 \\
 0 & 0 & 0 & 0 & 0 & 0 & i \omega  & -\frac{\Omega }{2} & 0 \\
 0 & 0 & 0 & 0 & 0 & 0 & \frac{\Omega }{2} & i \omega -\frac{\kappa }{2} & 0 \\
 0 & 0 & 0 & 0 & \kappa  & 0 & 0 & 0 & 0
\end{array}
\right)$
 \end{small}\\
 \begin{small}
 $F(t)=
\left(
\begin{array}{ccccccccc}
 \Lambda _+^2 & -\Lambda _+ \Lambda _0 & 0 & -\Lambda _+ \Lambda _0 & \Lambda _0^2 & 0 & 0 & 0 & 0
   \\
 \Lambda _+ \Lambda _0 & -\Lambda _- \Lambda _+ & 0 & -\Lambda _0^2 & \Lambda _- \Lambda _0 & 0 & 0
   & 0 & 0 \\
 0 & 0 & e^{-i t \omega } \Lambda _+ & 0 & 0 & -e^{-i t \omega } \Lambda _0 & 0 & 0 & 0 \\
 \Lambda _+ \Lambda _0 & -\Lambda _0^2 & 0 & -\Lambda _- \Lambda _+ & \Lambda _- \Lambda _0 & 0 & 0
   & 0 & 0 \\
 \Lambda _0^2 & -\Lambda _- \Lambda _0 & 0 & -\Lambda _- \Lambda _0 & \Lambda _-^2 & 0 & 0 & 0 & 0
   \\
 0 & 0 & e^{-i t \omega } \Lambda _0 & 0 & 0 & -e^{-i t \omega } \Lambda _- & 0 & 0 & 0 \\
 0 & 0 & 0 & 0 & 0 & 0 & e^{i t \omega } \Lambda _+ & -e^{i t \omega } \Lambda _0 & 0 \\
 0 & 0 & 0 & 0 & 0 & 0 & e^{i t \omega } \Lambda _0 & -e^{i t \omega } \Lambda _- & 0 \\
 \lambda _+ & 2 \gamma  \Lambda _0^2 & 0 & 2 \gamma  \Lambda _0^2 & \lambda _- & 0 & 0 & 0 & 1 \\
\end{array}
\right) $
 \end{small} \\


\section{Reference}
\begin{thebibliography}{10}

\bibitem{Schlosshauer} M. Schlosshauer. {\it Decoherence and the Quantum-to-Classical Transition} (Springer-Verlag, Berlin, 2007)
\bibitem{Nielsen} M. A. Nielsen and I. L. Chuang, {\it Quantum Computation and Quantum Information} (Cambridge University Press, Cambridge, 2000). 
\bibitem{Lidar} D. A. Lidar, I. L. Chuang, and K. B. Whaley. Phys. Rev. Lett. {\bf 81}, 2594 (1998)
\bibitem{Xue} P. Xue and Y.-F. Xiao. Phys. Rev. Lett. {\bf 97}, 140501 (2006)
\bibitem{Pechukas} P. Pechukas, Phys. Rev. Lett. {\bf 73}, 1060 (1994).
\bibitem{Li} K-H Li, Physics of open systems, Physics Reports, {\bf 134(1)}, 1–85 (1986)
\bibitem{Rivas} A. Rivas and S. F. Huelga, {\it Open Quantum Systems: An Introduction} (Springer, Berlin, 2012).
\bibitem{Einstein} E. Einstein, B. Podolsky, and N. Rosen, Phys. Rev. {\bf 47},  777 (1935).
\bibitem{Horodecki} R. Horodecki, P. Horodecki, M. Horodecki, and K. Horodecki, Rev. Mod. Phys. {\bf 81}, 865 (2009).
\bibitem{Ekert} A. K. Ekert, Phys. Rev. Lett. {\bf 67}, 661 (1991).
\bibitem{Bennett} C. H. Bennett, G. Brassard, C. Cr\'epeau, R. Jozsa, A. Peres, and W. K. Wootters, Phys. Rev. Lett. {\bf 70}, 1895 (1993).
\bibitem{Shor} P.W. Shor, Phys. Rev. A {\bf 52}, R2493 (1995).
\bibitem{Gottesman} D. Gottesman and I. Chuang, Nature (London) {\bf 402}, 390 (1999).
\bibitem{Lindblad} G. Lindblad, Commun. Math. Phys. {\bf 48}, 119 (1976). 
\bibitem{kraus}K. Kraus, {\it Fundamental Notes of Quantum Theory} (Springer-Verlag, Berlin, 1988). 
\bibitem{Dicandia} R. Di Candia, B. Mejia, H. Castillo, J. S. Pedernales, J. Casanova, and E. Solano. Phys. Rev. Lett. {\bf 111}, 240502 (2013).
\bibitem{Haroche} S. Haroche and J.-M. Raimond, {\it Exploring the Quantum} (Oxford University Press, Oxford, 2007). 
\bibitem{Rosario} A. Rosario, E. Massoni, and F. De Zela, Journal of Physics B: Atomic, Molecular and Optical Physics {\bf 45}, 095501 (2012).
\bibitem{Page} D. N. Page and W. K. Wootters, Phys. Rev. D 27, 2885 (1983).
\bibitem{Giovannetti} V. Giovannetti, S. Lloyd, and L. Maccone, Phys. Rev. D {\bf 92}, 045033 (2015)






\end{thebibliography}
\end{document}